\documentclass[pre,aps]{revtex4}
\usepackage{graphicx,epsf}
\usepackage{amssymb}
\usepackage{amsmath}
\usepackage{latexsym}

\def\be{\begin{equation}}
\def\ee{\end{equation}}
\def\bq{\begin{eqnarray}}
\def\eq{\end{eqnarray}}
\def\beq{\begin{eqnarray*}}
\def\eeq{\end{eqnarray*}}

\begin{document}

\title{MINIMAL UNIVERSAL MODEL FOR CHAOS IN LASER WITH FEEDBACK}

\author{Riccardo MEUCCI$^{1}$ \& Stefano EUZZOR$^{1}$, F. Tito ARECCHI$^{2}$, Jean-Marc GINOUX$^{3}$}

\affiliation{$^1$National Institute of Optics - CNR, Florence, Italy, riccardo.meucci@ino.it}

\affiliation{$^2$University of Florence, Florence, Italy}

\affiliation{$^3$Aix Marseille Univ, Universit\'{e} de Toulon, CNRS, CPT, Marseille, France}

\begin{abstract}
We revisit the model of the laser with feedback  and the minimal  nonlinearity  leading to chaos. Although the model has its origin in laser physics, with peculiarities related to the $CO_2$ laser, it belongs to the class of the three dimensional paradigmatic nonlinear oscillator models giving chaos.  The proposed model contains three key nonlinearities, two of which are of the type $xy$, where $x$ and $y$ are the fast and slow variables. The  third one is of the type $xz^2$, where $z$ is an intermediate feedback variable. We analytically demonstrate that it is essential for producing  chaos  via local or  global homoclinic bifurcations. Its electronic implementation in the range of kilo Hertz region  confirms its potential in describing phenomena  evolving  on different time scales.
\end{abstract}

\maketitle

\section{Introduction}
\noindent

Deterministic chaos has represented a crucial issue in laser physics because the Lorenz system [1963] is formally equivalent to laser equations \cite{haken1975}. The first evidence of chaos in lasers  was given in 1982 in a modulated single mode $CO_2$ laser \cite{PRL1982} and we had to wait 1986 for an evidence of Lorenz type chaos in  a particular and little used laser emitting in far infrared region (the so called class C-laser) \cite{PRL1986}. Commonly used single mode lasers are described by two rate equations and they are intrinsic stable devices (class B-laser). Feedback circuits are frequently used to improve  their stability properties by reducing  residual intensity and frequency fluctuations in order to match specific requests. However, a feedback can have the opposite effect, that is, enhancing the relaxation oscillations around the steady state solution. In other words, a simple linear filtering with the appropriate bandwidth on the laser output intensity can induce chaotic fluctuations on it \cite{PRA1986, PRL1987}. From the dynamical point of view, a feedback increases the dimensionality of the phase space from two to three whence chaos becomes possible.
In the last two decades of the past century, the use of $CO_2$ lasers for demonstrating such a behavior has had advantages, mainly due the convenient time scales associated with the laser intensity and the population inversion. The former, evolves on a time scale regulated by the parameter $k$ (the decay rate of photon number of the laser mode) imposed by the length of the optical cavity and its losses. In our case, $k$ is around $10^7$ $s^{-1}$. The latter, i.e. the decay rate of the population inversion, which is called $\gamma$. For a molecular laser as the $CO_2$ laser, $\gamma$ is of the order $10^3-10^4$ $s^{-1}$. These two parameters imply a resulting time scale given by

\[
\sqrt{k \gamma(p_0 - 1)},
\]

where $p_0$ is the pump strength normalized to the threshold value. Usually, $p_0$ is around 2. To be effective in producing chaos, the feedback loop should act on the above mentioned time scale, in other terms, we have the following condition to be satisfied:

\[
k > \beta > \gamma,
\]

where $\beta$ is the bandwidth of the feedback loop.
A simple three dimensional model accounting for the laser intensity $x$ with decay rate $k$, population inversion $y$ with decay rate $\gamma$ and feedback strength $z$ with decay rate $\beta$, explains qualitatively the observed dynamics but it does not yield an accurate matching with the experiment. The problem is overcome by introducing the so called 4-level model for the $CO_2$ laser. It consists in taking into account two resonant levels with population $N_1$ and $N_2$ and two rotational manifolds with population $M_1$ and $M_2$, respectively. In this refined model the dimensionality of the phase space is increased up to 6 instead of the previous 3. For an introduction to the 6 dimensional model and accurate numerical simulations on it, see for example Freire \textit{et al.} [2015]. To reduce the dimensionality two different approaches can be followed. The first one is to use a reduction based on the Center Manifold Theory (CMT) proposed by Varone \textit{et al.} [1995]. This analytical method implies a reduced four dimensional model with the addition of nonlinear terms whose physical interpretation is difficult to provide. A feasible reduction to three dimensions is imaginable considering that it has been obtained for the five dimensional model of the $CO_2$ laser with cavity losses modulation \cite{ciofini}. The second approach for an equivalent three dimensional model which is more physical and straightforward is to take advantage for the 4 level model only from the correct value of the laser intensity in the stationary regime. This condition implies the use of an artificial value of $\gamma$ which can be two orders of magnitude greater, that is, $10^4$ to $10^5$ $s^{-1}$. As the laser output intensity does not depend on $\beta$, we have to use an effective value of $\beta$ for the feedback variable $z$ up to $10^6$ $s^{-1}$. The advantage consists in keeping the original nonlinearity in the $x$ and $y$ differential equations given by their product $x y$. Considering the above advantages and in view of the extension to different dynamical systems ranging from neuron dynamics in the low frequency region (below 1 Hz) to high frequency domains (fast electronics, opto-electronics, etc.) we adopt the following three dimensional model:

\begin{equation}
\label{eq1}
\begin{aligned}
\dfrac{dx}{dt} & = - k_0 x \left( 1 + k_1 z^2 - y \right), \hfill \\
\dfrac{dy}{dt} & = - \gamma y - 2 \frac{k_0}{\alpha} x y + \gamma p_0, \hfill \\
\dfrac{dz}{dt} & = - \beta \left( z - B_0 + \frac{R}{\alpha} x \right), \hfill \\
\end{aligned}
\end{equation}

where $x$ is the fast variable (laser output intensity), $y$ is the slow variable (population inversion) and $z$ is the feedback variable affecting the fast one in a nonlinear way but regulated in linear way as the result of a low pass filter whose input is the fast variable summed to bias.
The parameter $\alpha$ is a suitable normalization of the fast variable $x$ which however does not alter its form considering that it is homogenous in $x$. If $\alpha = 2k_o / \gamma$, the adopted model is formally equivalent to the original model of the laser with feedback.
The paper is organized as it follows. First we introduce a numerical analysis followed by an experimental part containing an analog implementation of the oscillator. Second we compare the new model with other paradigmatic oscillators. Its potentialities are discussed in conclusions.

\section{Minimal Universal Model}

\subsection{Dimensionless form}

We propose the following change of variables and parameters to recast Eqs. (\ref{eq1}) in a dimensionless form. Let's pose:

\[
y \to p_0 y \mbox{, } t \to \frac{t}{\gamma} \mbox{, } \epsilon_1 = \frac{k_0}{\gamma} \mbox{, } \epsilon_2 = \frac{\beta}{\gamma} \mbox{, } B_1 = \frac{R}{\alpha} = \frac{R \gamma}{2 k_0}.
\]

Thus, the minimal universal model for chaos in laser reads:

\begin{equation}
\label{eq2}
\begin{aligned}
\dfrac{dx}{dt} & = - \epsilon_1 x \left( 1 + k_1 z^2 - p_0 y \right), \hfill \\
\dfrac{dy}{dt} & = - y - x y + 1, \hfill \\
\dfrac{dz}{dt} & =  - \epsilon_2 \left( z - B_0 + B_1 x \right), \hfill \\
\end{aligned}
\end{equation}

Let's notice that with the parameter set used in our experiment and analysis, $\epsilon_1 \gg 1$ and $\epsilon_2 \gg 1$. So, model (\ref{eq2}) is a \textit{slow-fast} dynamical systems involving two \textit{fast} times scales. In the following $B_0$ will play the role of a control parameter.

\subsection{Fixed points}

By using the classical nullclines method, it can be shown that the dynamical system (\ref{eq2}) admits four fixed points only two of which are positive.

\begin{equation}
\label{eq3}
\begin{aligned}
I_1 \left(0, 1, B_0 \right), \hspace{2cm} \\
I_2 \left(x^*, y^* = \frac{1}{1+x}, z^* = B_0 - B_1 x \right), \hfill \\
\end{aligned}
\end{equation}

where the expression of $x^*$ (too large to be explicitly written here since it comes from the solution of a cubic polynomial) depends on the control parameter $B_0$. In this problem all fixed points are supposed to be positive. Thus, starting from the right-hand-side of Eqs. (\ref{eq2}), it can be easily shown that:

\[
0 \leqslant x^* \leqslant p_0 - 1, \mbox{ } \frac{1}{p_0} \leqslant y^*,  \mbox{ } B_0 - B_1 \left( p_0 - 1 \right) \leqslant z^* \leqslant B_0.
\]

\subsection{Jacobian matrix}

The Jacobian matrix of dynamical system (\ref{eq2}) reads:

\begin{equation}
\label{eq4}
J = \begin{pmatrix}
(-1 + p_0 y - k_1 z^2) \epsilon_1 \  &  \  p_0 x \epsilon_1 \  &  \  -2 k_1 x z \epsilon_1 \vspace{6pt} \\
-y \  &  \  - 1 -x \  & \  0 \vspace{6pt} \\
- B_1 \epsilon_2 \  &  \  0 \  &  \  \epsilon_2 \vspace{6pt} \\
\end{pmatrix}
\end{equation}

By replacing the coordinate of the fixed points $I_1$ (\ref{eq3}) in the Jacobian matrix (\ref{eq4}) one obtains the Cayley-Hamilton third degree eigenpolynomial which reads:

\begin{equation}
\label{eq5}
\left( \lambda + 1 \right) \left( \lambda + \epsilon_2 \right) \left[ \lambda + \epsilon_1 \left(1 + B_0^2 k_1  - p_0 \right) \right] = 0
\end{equation}

Thus, provided that:

\begin{equation}
\label{eq6}
p_0 < 1 + B_0^2 k_1,
\end{equation}

the fixed point $I_1$ is a \textit{saddle node}. Moreover, such condition (\ref{eq6}) provides a upper boundary for the control parameter $B_0$:

\begin{equation}
\label{eq7}
B_0 < \sqrt{\frac{p_0 - 1}{k_1}}.
\end{equation}

From the positivity of the fixed points, let's notice that if $B_0 = B_1(p_0 - 1)$, then the second fixed point $I_2$ reads: $I_2 (x^* = p_0 -1, y^* = 1/p_0, z^* = 0)$. In these conditions and according to Eq. (\ref{eq7}), it can be stated that:

\begin{equation}
\label{eq8}
B_1 \left( p_0 - 1 \right) < B_0 < \sqrt{\frac{p_0 - 1}{k_1}}.
\end{equation}

Thus, for $B_0 = B_1(p_0 - 1)$, computation of the eigenvalues of $I_1$ shows that two of them are real and negative and one is real and positive confirming thus the \textit{saddle node} feature of this point while for $I_2$, one is real and negative and the two others are complex conjugate with negative real parts. So, in this case the fixed point $I_2$ is stable and attractive according to Lyapunov. For $B_0 = \sqrt{(p_0 - 1)/k_1}$, we found that both fixed points $I_1$ and $I_2$ are stable and attractive (all the real parts of their eigenvalues are negative). Such a result will enable to explain the limits of the bifurcation diagram presented below (see Fig. 1 \& 2) outside which no attractor can exist. Then, while using the parameters set of our experiment, i.e., for any value of $B_0 \in [B_1 ( p_0 - 1 ), \sqrt{(p_0 - 1)/k_1}]$ it can be shown that $I_2$ is a \textit{saddle-focus} (the first eigenvalue is real and negative while the two others are complex conjugate with positive real parts). It implies that a Hopf bifurcation occurs with this interval \cite{Hopf, Andronov, Marsden, Kuznetsov}. Nevertheless, the fact that the fixed points are solutions of a cubic polynomial precludes from explicit the value of the control parameter $B_0$ for which such a Hopf bifurcation occurs. However, we have found a method (see Appendix) allowing to analytically compute the upper bound of such a parameter. In the following we will use the parameters set in our experiment and analysis:

\[
\epsilon_1 = 200 \mbox{ , } \epsilon_2 = 6 \mbox{ , } k_1 = 12 \mbox{ , } p_0 = 1.208  \mbox{ , } B_1 = 0.555.
\]

\subsection{Bifurcation diagram}

Thus, in order to highlight the effects of the control parameter $B_0$ changes on the topology of the attractor we have  built a bifurcation diagram (see Fig. 1 \& 2) that we have compared to the phase portraits plotted in Fig. 3. First, we observe that for $B_0 \approx 0.12$ a Hopf bifurcation occurs (see Appendix). Then, for $B_0 = 0.123$, a \textit{limit cycle} appears. As $B_0$ increases between $0.123$ and $0.1237$, a ``period doubling cascade'' occurs and so, we observe a \textit{2-periodic limit cycle}. In the interval $0.1237 < B_0 < 0.12425$, the period of the \textit{limit cycle} increases again and becomes equal to the number of branches in the bifurcation diagram (see Fig. 2 and Fig. 3a). For $0.12425 < B_0 < 0.129$, a \textit{stable homoclinic orbit} appears and persists (see Fig. 2d and Fig. 3b).

\begin{figure}[htbp]
\centerline{\includegraphics[width=8cm,height=8cm]{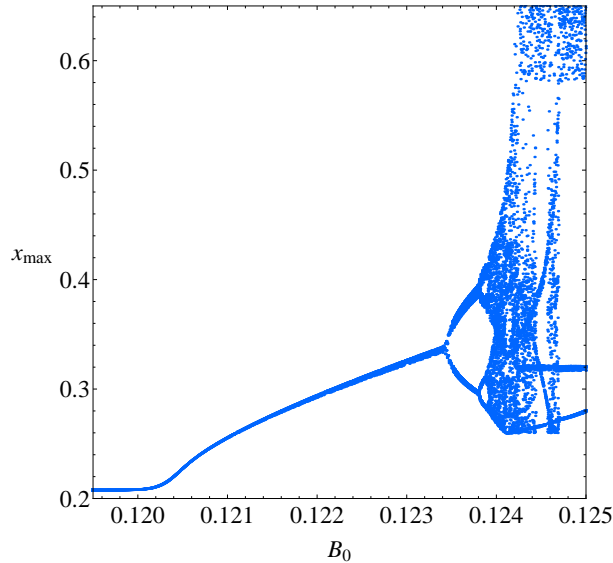}}
\caption{Bifurcation diagram $x_{max}$ as function of $B_0$.}
\label{fig1}
\end{figure}

\begin{figure}[htbp]
\centerline{\includegraphics[width=8cm,height=8cm]{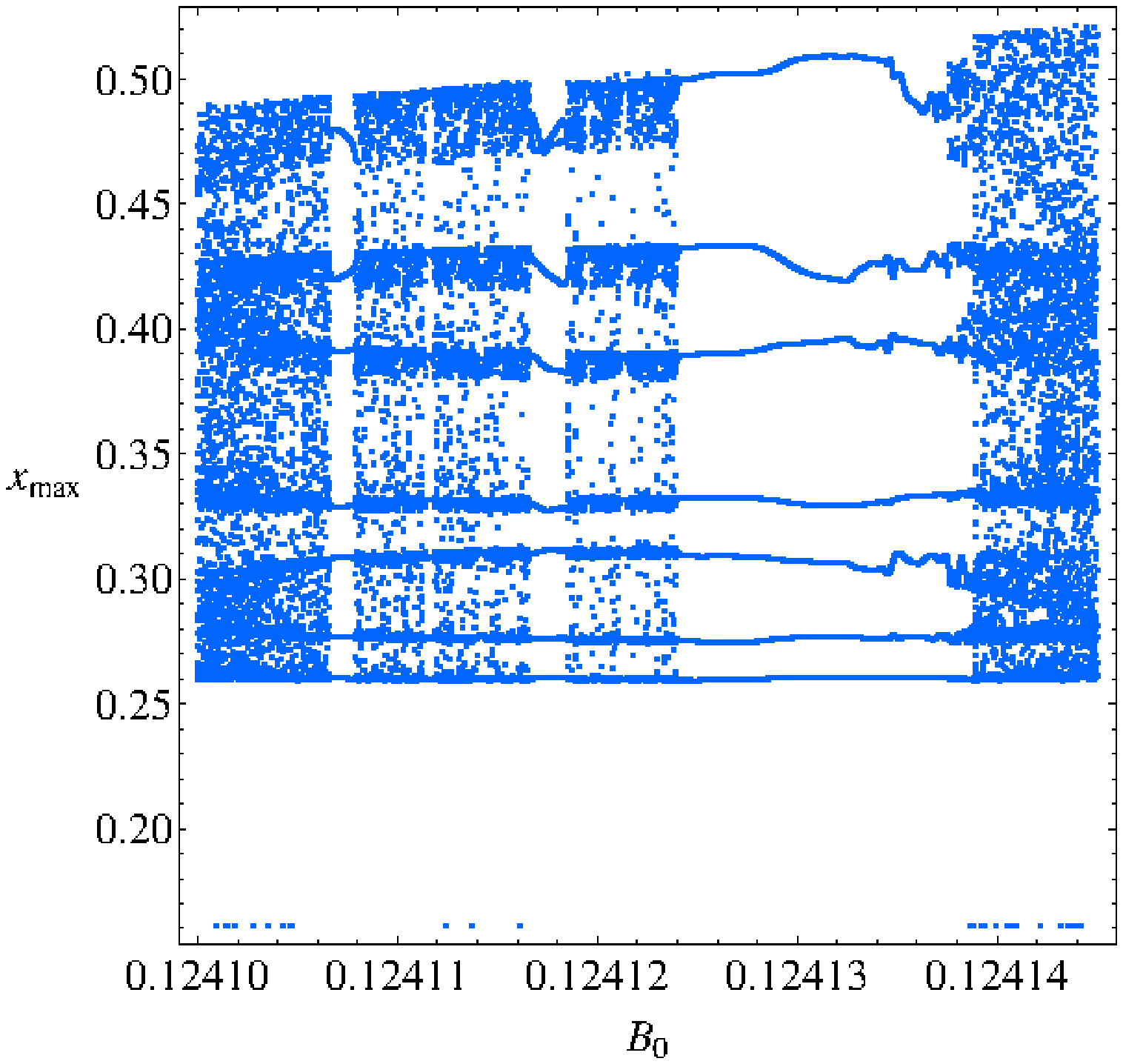}}
\caption{zoom of the bifurcation diagram $x_{max}$ as function of $B_0$.}
\label{fig2}
\end{figure}

\begin{figure}[htbp]
\centerline{\includegraphics[width=6cm,height=6cm]{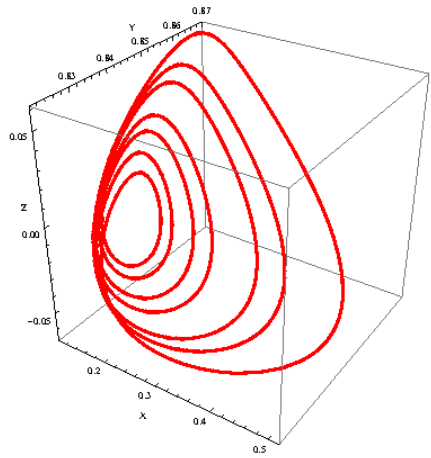}}
\centerline{(a) $B_0 = 0.12413$} \vspace{0.1in}
\centerline{\includegraphics[width=6cm,height=6cm]{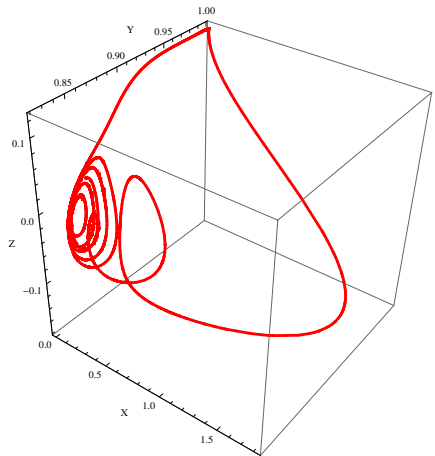}}
\centerline{(b) $B_0 = 0.1243$} \vspace{0.1in}
\centerline{\includegraphics[width=6cm,height=6cm]{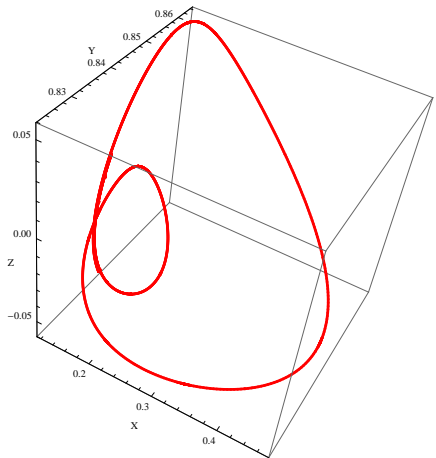}}
\centerline{(c) $B_0 = 0.12463$} \vspace{0.1in}
\caption{Phase portraits of model (\ref{eq2}) for various values $B_0$.}
\label{fig3}
\end{figure}

\begin{figure}[htbp]
\centerline{\includegraphics[width=6cm,height=6cm]{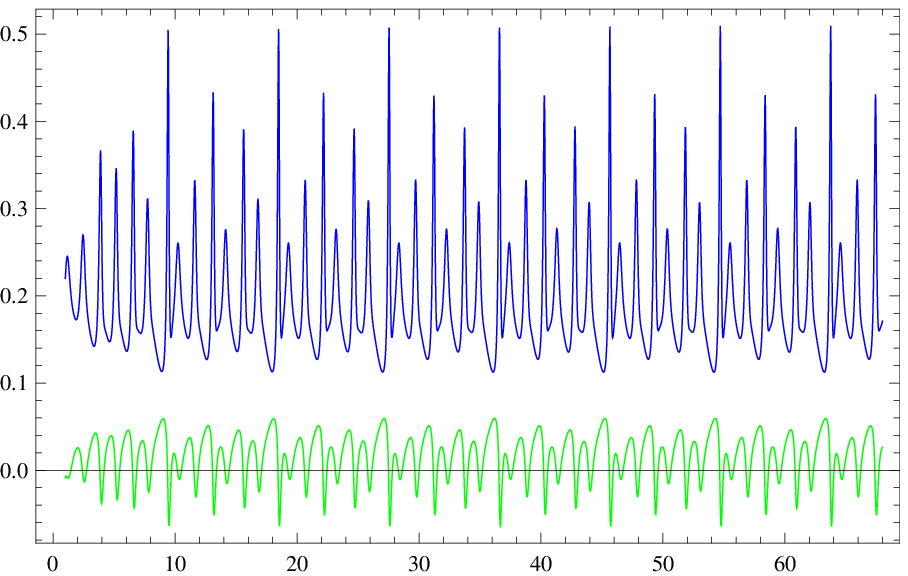}}
\centerline{(a) Time series $x(t)$ and $z(t)$ for $B_0 = 0.12413$} \vspace{0.1in}
\centerline{\includegraphics[width=6cm,height=6cm]{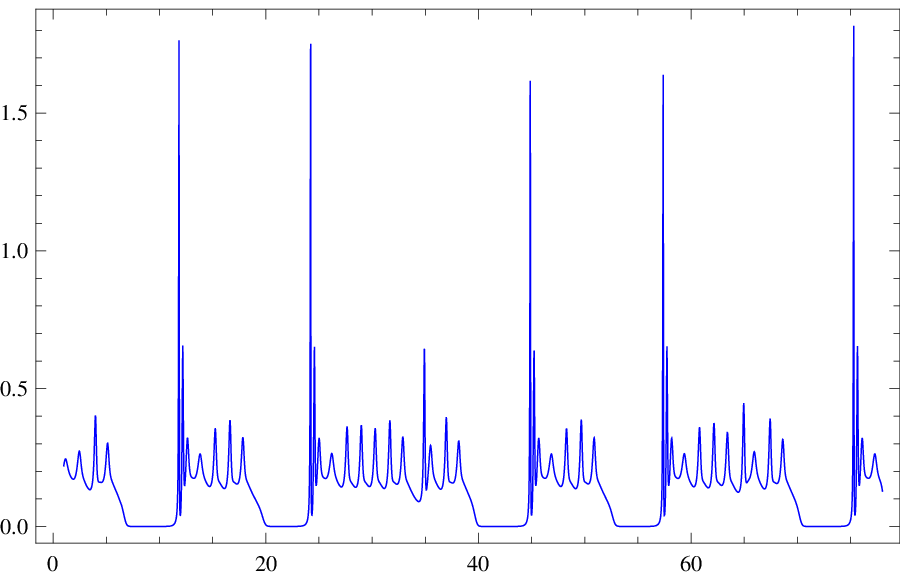}}
\centerline{(b) Time series $x(t)$ for $B_0 = 0.1243$} \vspace{0.1in}
\centerline{\includegraphics[width=6cm,height=6cm]{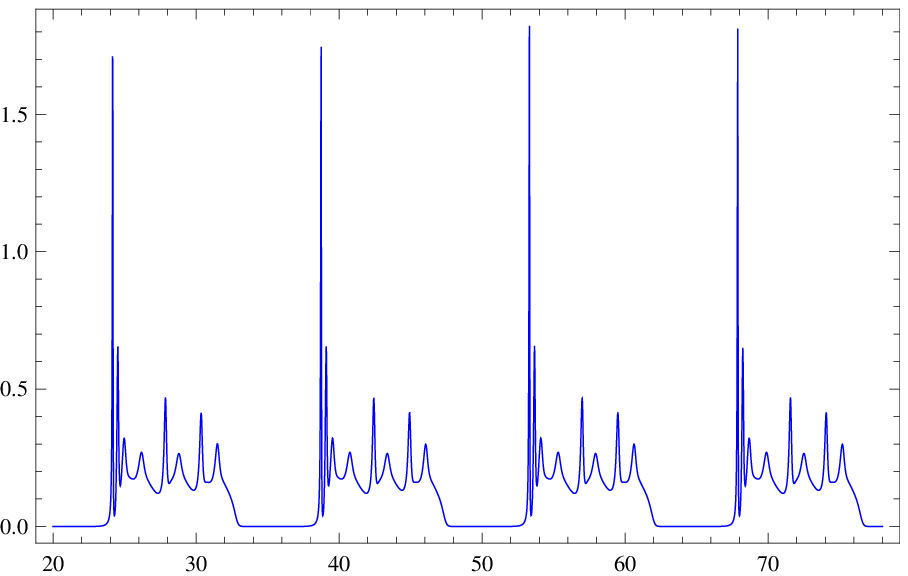}}
\centerline{(c) Time series $x(t)$ for $B_0 = 0.12463$} \vspace{0.1in}
\caption{Time series of model (\ref{eq2}) for various values $B_0$.}
\label{fig4}
\end{figure}

In order to confirm such scenario, Lyapunov Characteristic Exponents (LCE) have been computed in each case.

\subsection{Numerical computation of the Lyapunov exponents}

The algorithm developed by Marco Sandri [1996] for Mathematica$^{\mbox{\scriptsize{\textregistered}}}$ has been used to perform the numerical calculation of the Lyapunov characteristics exponents (LCE) of dynamical system (\ref{eq2}) in each case. LCEs values have been computed within each considered interval ($B_0 \in [0.123, 0.1234]$ and $[0.1235, 0.125]$). As an example, for $B_0 = 0.123, 0.124$ and $0.1246$, Sandri's algorithm has provided respectively the following LCEs $(0, -0.56, -6.63)$, $(+0.27, 0, -7.47)$ and $(+0.2, 0, -7.32)$. Then, following the works of Klein and Baier [1991], a classification of (autonomous) continuous-time attractors of dynamical system (\ref{eq2}) on the basis of their Lyapunov spectrum is presented in Tab. 1. LCEs values have been also computed with the Lyapunov Exponents Toolbox (LET) developed by Steve Siu for MatLab$^{\mbox{\scriptsize{\textregistered}}}$ and involving the two algorithms proposed by Wolf \textit{et al.} [1985] and Eckmann and Ruelle [1985] (see https://fr.mathworks.com/matlabcentral/fileexchange/233-let). Results obtained by both algorithms are consistent.

\begin{table}[h]
\centering
\caption{Lyapunov characteristics exponents of dynamical system (\ref{eq2}) for various values of $B_0$.}
{\begin{tabular}{c c c}\\[-2pt]
{\hspace{11mm} $m$} & LCE spectrum & Dynamics of the attractor  \\[6pt]
\hline\\[-2pt]
{$0.1230 < B_0 < 0.1234$} & ($ 0, -, - $) & Periodic Motion \\[1pt]
{$0.1235 < B_0 < 0.124$} & ($ 0, -, - $) & $n$-Periodic Motion  \\[2pt]
{$0.1241 < B_0 < 0.125$} & ($ +, 0, - $) & Homolinic Chaos  \\[2pt]
\end{tabular}}
\label{tab1}
\end{table}

\section{Experimental part}

The  set up used in our experiment is illustrated in Fig. 7. Briefly, it consists of three integrators $I_1$, $I_2$ and $I_3$ (LT1114 by Analog Devices), whose outputs are the signals $x$, $y$ and $z$ and $y = \dot{x}$ contained in Eqs. \ref{eq1}. The other integrator $I_4$, is employed for an inverting amplifier with unitary gain. The two nonlinearities are implemented by means of three analog multipliers $M_1$, $M_2$ and $M_3$ (MLT04, by Analog Devices). The first one yields the product $x y$, while the other two multipliers implement the product $x z^2$. The semplification of the proposed scheme is evident when compared with the one obtained using a Field  Programmable Analog  Array (FPAA) circuit\cite{fortunafrasca2005}.

\begin{figure}[htbp]
\centerline{\includegraphics[width=6cm,height=6cm]{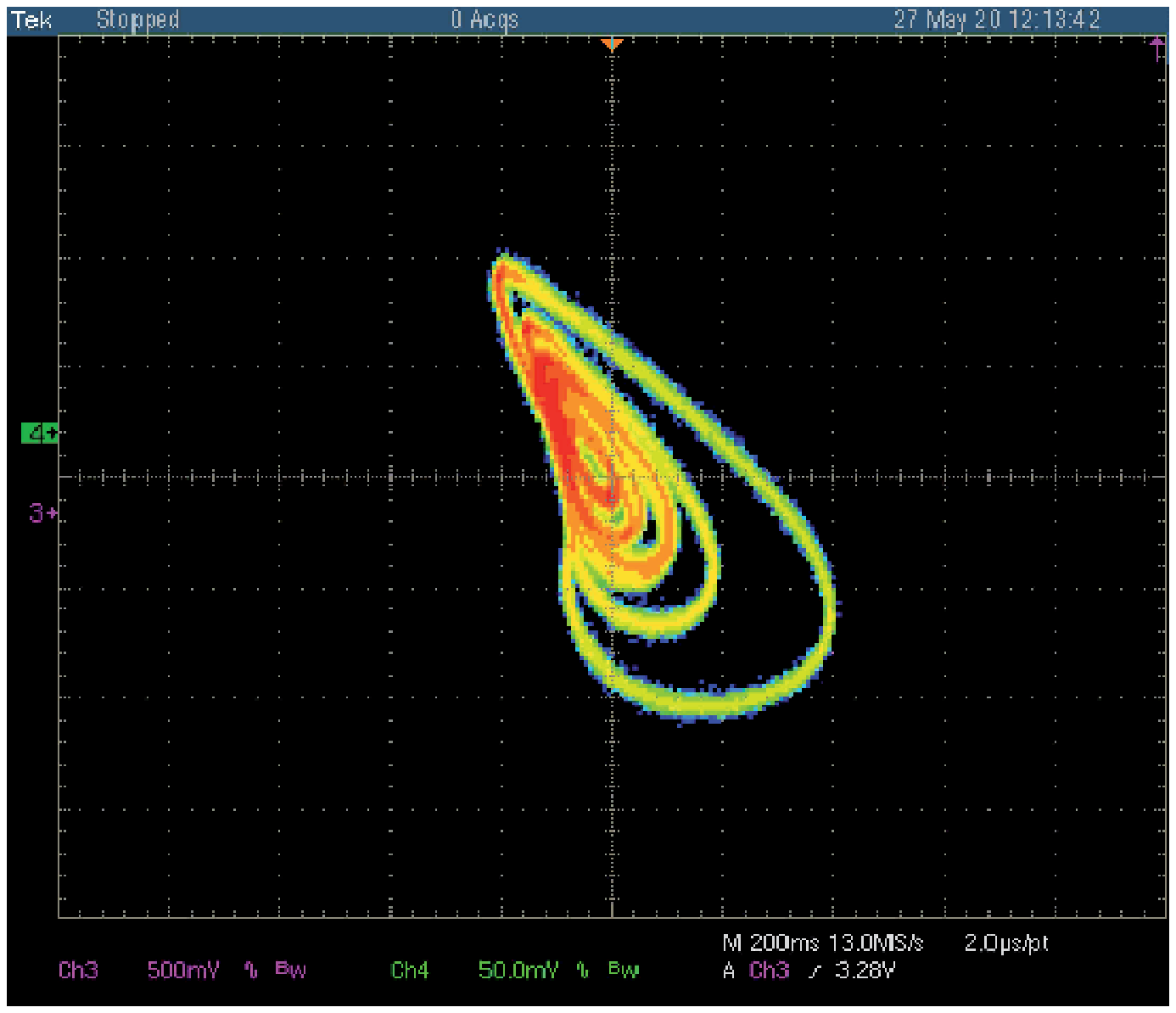}}
\vspace{0.1in}
\centerline{(a) Phase portrait for $V(B_0) = -2.550V$}
\vspace{0.1in}
\centerline{\includegraphics[width=6cm,height=6cm]{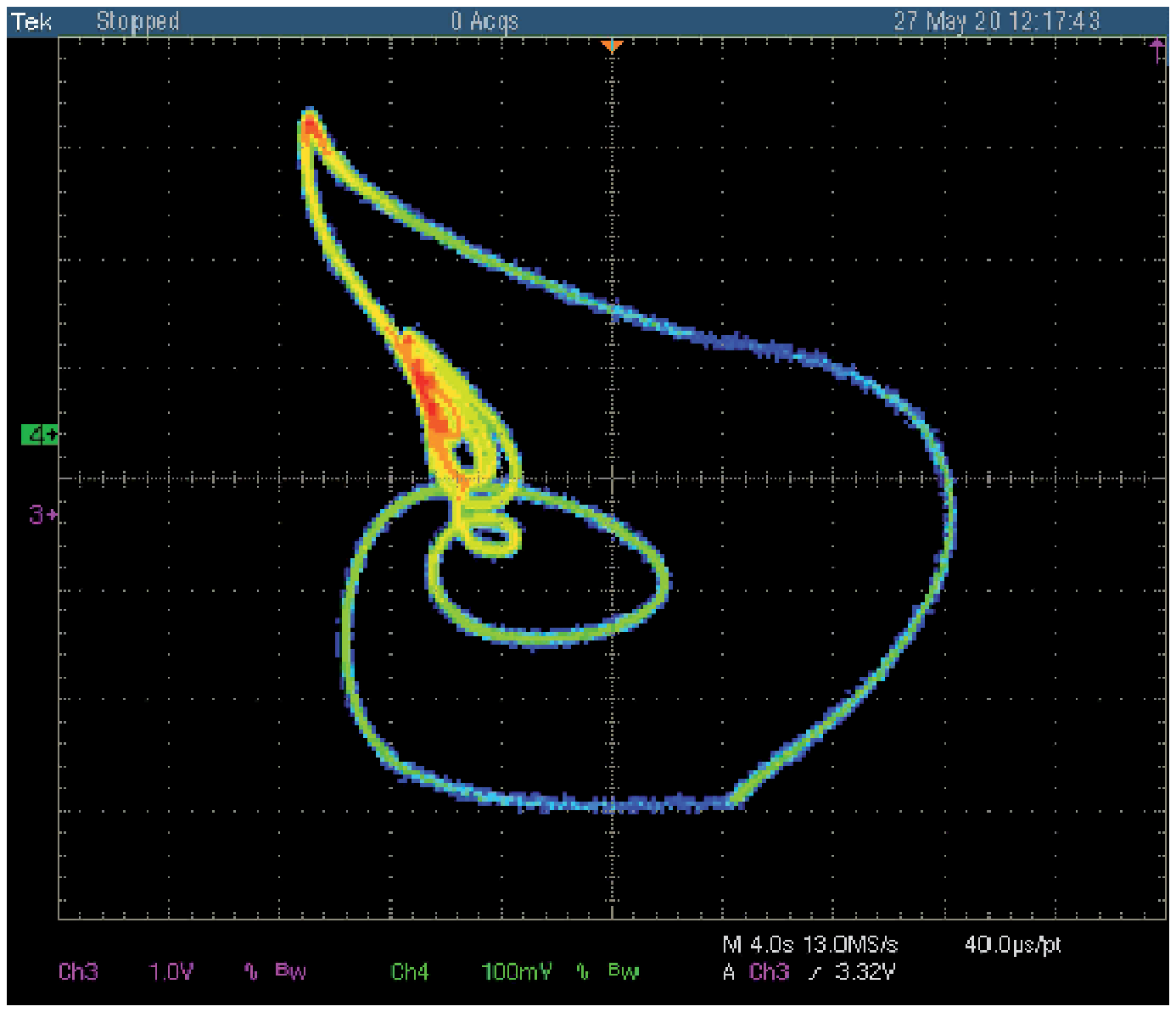}}
\vspace{0.1in}
\centerline{(b) Phase portrait for $V(B_0) = -2.596V$}
\vspace{0.1in}
\centerline{\includegraphics[width=6cm,height=6cm]{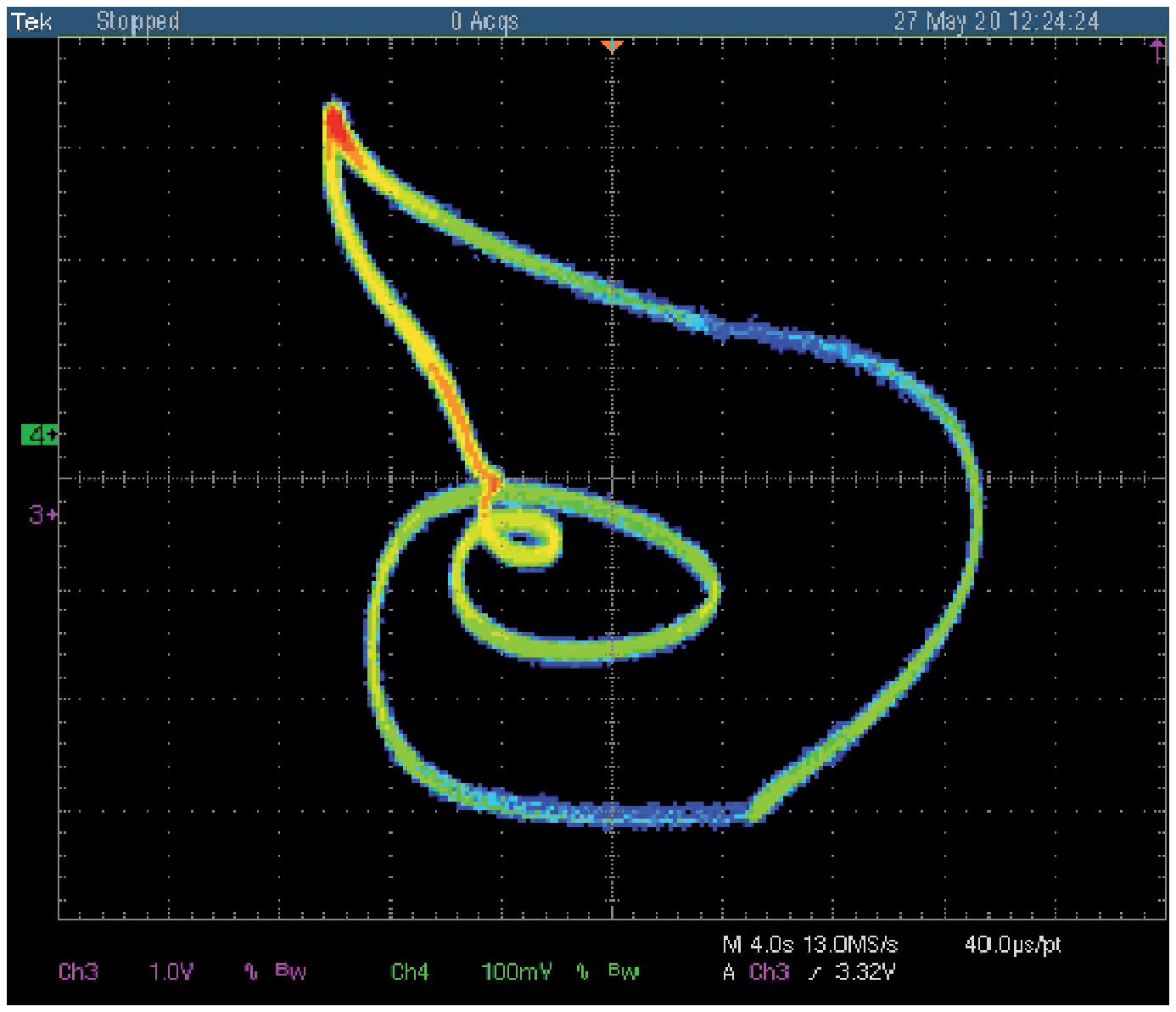}}
\vspace{0.1in}
\centerline{(c) Phase portrait for $V(B_0) = -2.628V$}
\vspace{0.1in}
\caption{Oscilloscope snapshots of phase portraits for $V(p_0) = -2.983 V$ and various values of $V(B_0)$.}
\label{fig5}
\end{figure}

\begin{figure}[htbp]
\centerline{\includegraphics[width=6cm,height=6cm]{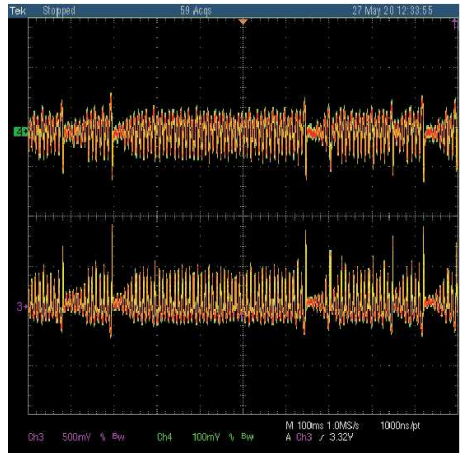}}
\vspace{0.1in}
\centerline{(a) Time series $x(t)$ and $z(t)$ for $V(B_0) = -2.550V$}
\vspace{0.1in}
\centerline{\includegraphics[width=6cm,height=6cm]{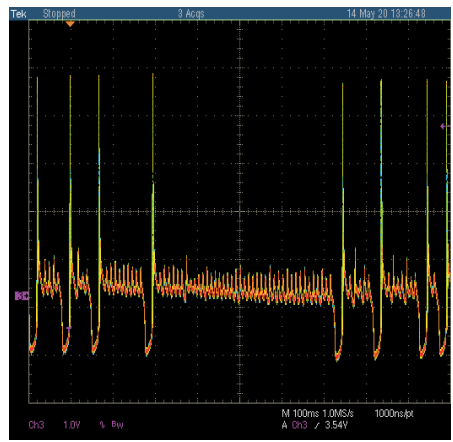}}
\vspace{0.1in}
\centerline{(b) Time series $x(t)$ for $V(B_0) = -2.596V$}
\vspace{0.1in}
\centerline{\includegraphics[width=6cm,height=6cm]{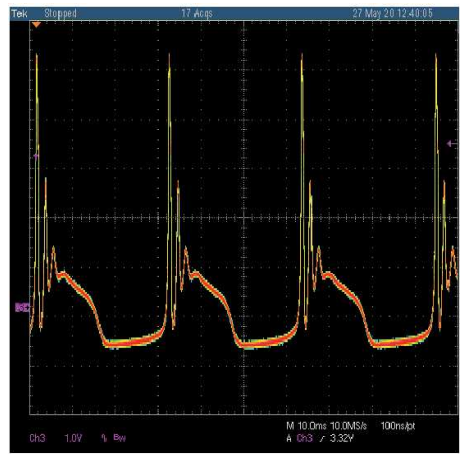}}
\vspace{0.1in}
\centerline{(c) Times series $x(t)$ for $V(B_0) = -2.628V$}
\vspace{0.1in}
\caption{Oscilloscope snapshots of time series for $V(p_0) = -2.983 V$ and various values of $V(B_0)$.}
\label{fig6}
\end{figure}

\begin{figure}[htbp]
\centerline{\includegraphics[width=9 cm,height=11cm]{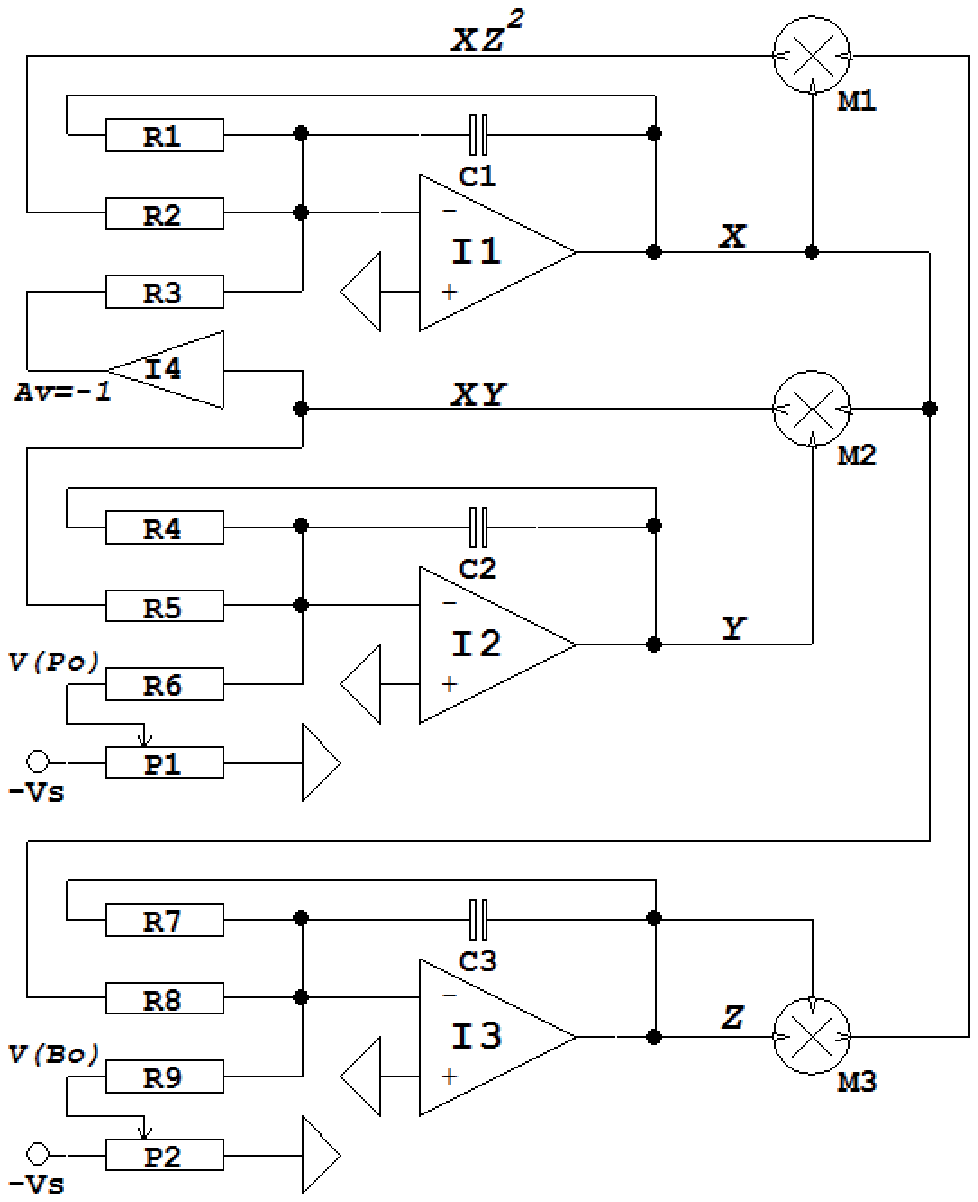}}
\caption{Circuit diagram of the ``laser with feedback''.}
\label{fig7}
\end{figure}

Considering the limits imposed by analog simulations, the  desired dynamics of the oscillator is obtained by fine adjustments of  two bias voltages $V(p_0)$ and $V(B_0)$ accounting for the  parameters, $p_0$ and $B_0$  in Eqs. \ref{eq1}. This condition  has been achieved by means of two potentiometers $P_1$ and $P_2$ connected to a fixed negative voltage source $-Vs$. The relaxation rates of the three variables (reciprocal of the integration times of the three integrators), according to Eqs. \ref{eq1}  and after a temporal rescaling of three orders of magnitude, are selected as follows. For the $x$ integrator we set  $k_0 = 1/(R_1 C_1)= 2.463 \times 10^4s^{-1}$ where $R_1 = 33 k \Omega$ and $C_1 = 1.67nF$, ($R_3 = R_1 = 33 k \Omega$). For the $y$  integrator we set  $\gamma = 1/ (R_4 \times C_2) = 1.00 \times 10^2s^{-1}$, where $R_4 = 10 k \Omega$, $C_2 = 1 \mu F$. For the $z$ integrator we select  $\beta = 1/ (R_7 \times C_3 ) = 1.00 \times 10^3s^{-1}$, with $R_7 = 10 k \Omega$ and $C_3 = 100nF$. For the other parameter values we have:$ k_1 = R_1/(R_2 \times 2.5) = 13.2$ where $R_2 = 1 k \Omega$; $p_0 = R_4/R_6 = 1.196$ where $R_6=8.36 k \Omega$;  $B_0 = R_7/R_9 = 0.115$ where $R_9 = 86.6 k \Omega$; $R/\alpha = R_7/R_8 = 0.222$, where $R_8 = 45 k \Omega$.

The attractors in the $x-z$ phase space for different values of the $B_0$ parameter value are reported in Figs. 5.  In panel a) the local dynamics emerging after an Hopf bifurcation is shown. An increase of the control parameter $B_0$ leads to a condition of homoclinic chaos due to a homoclinic orbit around which a chaotic regime characterized by pulses nearly of the same high but erratically separated in time due to the rejection mechanism around the local chaos shown in a). A successive increment of the control parameter leads to the stabilization of the pulsed regime which are typical in a relaxation oscillator. In the presented attractors (see b and c) we observe a small distortion for high amplitudes of the $x$ signal due to the frequency limitation of the analog multipliers. This limitation does not induce severe limitations to the global dynamics as shown in the temporal behavior of the $x$ signal, when we consider applications in the low frequency range below 1 kHz. For applications in high frequency regimes the nonlinearities must be implemented by using CMOS devices (range 1-10 MHz).

\section{Discussion}

Numerical analysis of the proposed model and its electronic implementation confirm its potentialities that promote it to the class of  as other well-known paradigmatic models as the Lorenz's \cite{lorenz1963}, Chua's \cite{matsumoto1984,chua1986} Chen's \cite{guanrong2010,chen2002},  Roessler's \cite{roessler1976} and other simple three dimensional systems reported by Sprott \cite{sprott}. The Lorenz model is formally equivalent to the laser equations (the so called class C-laser), as demonstrated by H. Haken \cite{haken1975} so the link with laser dynamics is direct. The intrinsic symmetries imply a chaotic trajectory visiting two saddle foci to be contrasted with the laser with feedback where the competing steady states are a saddle focus and a saddle node which contributes to the creation of the homoclinic connection. The comparison with the Chua's circuit which relies on a ``locally active resistor'' with static nonlinear characteristic is of another kind. This element is the Chua's diode and it can be implemented in different ways. The chaotic attractors from the Chua circuit perfectly reproduce Lorenz chaos even though the correspondence with laser equations is difficult to draw. The chaotic Chen system is similar but not equivalent to the Lorenz one as recently pointed out by Chen \cite{chen2020} introducing of the concept ``generalized Lorenz systems family''. In both cases the nonlinearities are two of the quadratic type in models which have seven terms on the right hand side. The Roessler circuit is simpler when compared with Lorenz systems because it possesses only one quadratic term and seven terms on the right hand side. It does not seem that both Chua and Roessler systems are related to laser dynamics. The above cited systems played a crucial role in chaos synchronization demonstrated by Pecora  and Carroll \cite{pecoracarroll1990} using Lorenz and Roessler circuits.
The increased complexity in the laser with feedback is due to feedback process which implies the additional cubic term $xz^2$ to the two quadratic nonlinearities $xy$. This is the first time that the cubic term is treated in its simplest form approximating $\sin(z)^2$ with $z^2$. It is important to stress that adiabatic elimination of the fast variable corresponding to laser polarization in the Lorenz system does not lead to rate model of the laser equations the so called class B laser, the large class of the available lasers, including semiconductor lasers and solid state lasers. As far as laser semiconductor lasers are concerned, it is important to note that their dynamics is well described by the Lang and Kobayashi  (LK) model \cite{Lang} accounting for the effects of delayed optical feedback acting on the timescale of the intrinsic semiconductor laser. The LK equations  describe the complex  dynamics of the complex electric field $E$ and the inversion (number of electron-hole pairs) $N$ inside the laser. The fast chaotic dynamics from these laser has been largely used in applications  for secure communication systems (\cite{Fischer, roy, donati, Ohtsubo}).

\section{Conclusions}

In conclusion, we retain that the laser with feedback, whose historical origins dates back in the same period of other chaotic oscillators will be appropriate for describing instances of local chaos, reached after subharmonic bifurcations, global bifurcations as homoclinic chaos and relaxation type oscillation behavior or regular spiking behavior when a control parameter is changed. Another valuable advantage over the nonlinear oscillators described above is related to the fact that  elimination of the feedback variable leaves unchanged its potentialities and chaos can be reached by modulation of the parameter $k$ provided the general condition on the timescales fulfilled. In this framework, interesting perspectives are also foreseen for competition population dynamics ruled by  Volterra-Lotka models \cite{volterra1926, volterra1931, lotka1910, lotka1920} and  chaotic epidemiological models \cite{SEIR}, in both cases the underlying nonlinearities are of the direct product of two variables.

\section{Appendix}

This appendix presents a method allowing to provide a upper bound for the Hopf bifurcation parameter for any three-dimensional autonomous dynamical system for which the fixed points coordinates cannot be easily expressed analytically as it is the case for the dynamical system (\ref{eq2}) for which the coordinates of the fixed point $I_2$ are the roots of a cubic polynomial. Let's suppose that the three eigenvalues of the Jacobian matrix $J$ of this dynamical system evaluated at the fixed point ($I_2$ in our case) are real and complex conjugate $\lambda_1$, $\lambda_{2, 3} = \alpha \pm i \omega$. The Cayley-Hamilton eigenpolynomial reads:

\begin{equation}
\label{eq9}
\lambda^3 -\sigma_1 \lambda^2 + \sigma_2 \lambda - \sigma_3 = 0
\end{equation}

where $\sigma_1 = Tr \left( J \right)$, $\sigma_2 = \sum_{i = 1}^3 M_{ii} \left( J \right)$ is the sum of all first-order diagonal minors of $J$ and $\sigma_3 = Det \left( J \right)$. Thus, we have:

\begin{equation}
\label{eq10}
\begin{aligned}
\sigma_1 & = Tr \left( J \right) = \lambda_1 + \lambda_2 + \lambda_3 = \lambda_1 + 2 \alpha, \hfill \\
\sigma_2 & = \sum_{i = 1}^3 M_{ii}  \left( J \right) = \lambda_1 \lambda_2 + \lambda_1 \lambda_3 + \lambda_2 \lambda_3 = 2 \alpha \lambda_1 + \beta, \hfill \\
\sigma_3 & = Det \left( J \right) = \lambda_1 \lambda_2 \lambda_3 = \lambda_1 \beta, \hfill \\
\end{aligned}
\end{equation}

where $\beta = \alpha^2 + \omega^2$. In order to analyze the stability of fixed points according to a control parameter value ($B_0$ here), i.e. the occurrence of Hopf bifurcation, we propose to use the Routh-Hurwitz' theorem \cite{Routh1877,Hurwitz1893} which states that if $D_1 = \sigma_2$ and $D_2 = \sigma_3 - \sigma_2 \sigma_1$ are both positive then \textit{eigenpolynomial} equation (\ref{eq9}) would have eigenvalues with negative real parts. From Eqs. (\ref{eq10}) it can be stated that:

\begin{equation}
\label{eq11}
\alpha = \frac{\sigma_1 \sigma_2 - \sigma_3}{\lambda_1^2 + \sigma_2}
\end{equation}

Thus, $\alpha = 0$ provided that $\kappa = \sigma_1 \sigma_2 - \sigma_3 = 0$. For dynamical system (\ref{eq2}), we obtain:

\begin{equation}
\label{eq12}
\kappa = \left( 1 + x \right) \left[ p_0 x y \epsilon_1 + \left( 1 + x + \epsilon_2 \right) \epsilon_2 \right] -2 B_1 k_1 x z \epsilon_1 \epsilon_2^2
\end{equation}

then, by replacing $x$, $y$ and $z$ the coordinates (\ref{eq3}) of the fixed point $I_2$, we have:

\begin{equation}
\label{eq13}
B_0 = \frac{ 1 + \epsilon_2}{2 B_1 k_1 x^* \epsilon_1 \epsilon_2} + \frac{p_0 \epsilon_1 + \epsilon_2  (2 + \epsilon_2)}{2 B_1 k_1 \epsilon_1 \epsilon_2^2} + \frac{(1 + 2 B_1^2 k_1 \epsilon_1 \epsilon_2)x^*}{2 B_1 k_1 \epsilon_1 \epsilon_2}
\end{equation}

Positivity of fixed points has led to $x^* \leqslant p_0 -1$ which implies that $\max\left( x^* \right) = p_0 - 1$. Thus, by posing $x^* = p_0 - 1$ in (\ref{eq13}) and while using the parameters sets of our experiment, i.e., $\epsilon_1 = 200$, $\epsilon_2 = 6$, $k_1 = 12$, $p_0 = 1.208$ and $B_1 = 0.555$, we find:

\[
B_0^{Hopf} \leqslant 0.12057
\]

The numerical computation of the Hopf bifurcation parameter value has been found equal to $0.12036$ which is below and very near the upper bound analytically obtained.


\begin{thebibliography}{9}


\bibitem[Andronov {\it et al.}(1971)]{Andronov}
Andronov, A., Leontovich, E., Gordon, I. \& Maier, A. [1971], {\it Theory of Bifurcations of Dynamical Systems on a Plane}, Israel Program for Scientific Translations, Jerusalem.

\bibitem[Arecchi {\it et al.}(1982)]{PRL1982}
Arecchi, F. T.,  Meucci, R.,  Puccioni, G. P. \&  Tredicce, J. R. [1982] ``Experimental evidence of subharmonic bifurcations, multistability, and turbulence in a
q-switched gas laser,'' \emph{Phys. Rev. Lett.} 49, 1217-1220.

\bibitem[Arecchi {\it et al.}(1986)]{PRA1986}
Arecchi, F. T. , Gadomski, W. \& Meucci, R. [1986] ``Generation of chaotic dynamics by feedback on a laser,'' \emph{Phys. Rev.A.} 34, 1617-1620.

\bibitem[Arecchi {\it et al.}(1987)]{PRL1987}
Arecchi, F. T., Meucci, R. \& Gadomski, w. [1987] ``Laser dynamics with competing instabilities,'' \emph{Phys. Rev. Lett.}, 58, 2205-2208.

\bibitem[Arecchi {\it et al.}(2005)]{fortunafrasca2005}
Arecchi, F. T., Fortuna, L., Frasca, M., Meucci, R. \& Sciuto, G. [2005] ``A programmable electronic circuit for modelling $CO_2$ laser dynamics,'' \emph{Chaos} \textbf{15}, 043104.

\bibitem[Celikovshy \& Chen(2002)]{chen2002}
Celikovshy, S. \& Chen, G. [2002] ``On a generalized Lorenz canonical form of chaotic systems,'' \emph{Int. J. Bifurcation Chaos}, 12, 1789-1812.

\bibitem[Chen(2020)]{chen2020}
Chen, G.  [2020] ``Generalized Lorenz systems family,'' arXiv:2006.04066.

\bibitem[Chua {\it et al.}(1986)]{chua1986}
Chua, L.O., Kumaro, M.  \& Matsumoto, T. [1986] ``The double scroll family,'' \emph{IEEE Transactions on Circuits and Systems}, 33, 1072-1118.

\bibitem[Ciofini {\it et al.}(1993)]{ciofini}
Ciofini, M., Politi, A. \& Meucci, R. [1993] ``Effective two-dimensional model for $CO_2$ lasers,'' \emph{Phys. Rev. A} 48, 605-610.

\bibitem[Donati \& Mirasso(2002)]{donati}
Donati, S.  \&  Mirasso, C. R. [2002] ``Feature section on optical chaos and applications to cryptography,'' \emph{IEEE J. Quantum Electron.} 38, 1138-1205.

\bibitem[Eckmann \& Ruelle(1985)]{Eckmann1985}
Eckmann, J. P.  \& Ruelle, D. [1985] ``Ergodic theory of chaos and strange attractors,'' \emph{Rev. Mod. Phys.}, \textbf{57}, 617-656.

\bibitem[Fischer \textit{et al.}(2000)]{Fischer}
Fischer, I., Liu, Y. \& Davis, P. [2000] ``Synchronization of chaotic semiconductor laser dynamics on subnanosecond time scales and its potential for chaos communication,'' \emph{Phys. Rev. A} 62, 011801.

\bibitem[Freire {\it et al.}(2015)]{freire2015}
Freire, J. G., Meucci, R., Arecchi, F. T. \&  Gallas, J. A. C. [2015] ``Self-organization of pulsing and bursting in a $CO_2$ laser with opto-electronic feedback,'' \emph{Chaos}, 25(9):097607.

\bibitem[Haken(1975)]{haken1975}
Haken, H. [1975] ``Analogy between higher instabilities in fluids and lasers,'' \emph{Phys. Lett.A}, 53, 77-78.

\bibitem[Hau {\it et al.}(2010)]{guanrong2010}
Hau, Z., Kang, N., Kong, X., Chen, G. \& Yan, G. [2010] ``On the equivalence of Lorenz system and Chen system,'' \emph{Int. J. Bifurcation Chaos}, 20, 557-560.

\bibitem[Hopf(1942)]{Hopf}
Hopf, E. [1942] ``Abzweigung einer periodischen L\"osung von einer station\"aren L\"osung eines Differentialsystems,'' \emph{Berichte der MathematischPhysikalischen Klasse der S\"achsischen Akademie der Wissenschaften zu Leipzig}, Band XCIV, Sitzung vom 19. Januar 1942, pp. 3-22. See L. N. Howard and N. Kopell, A Translation of Hopf's Original Paper, pp. 163-193 and Editorial Comments, pp. 194-205 in J. Marsden and M. McCracken.

\bibitem[Hurwitz(1893)]{Hurwitz1893} Hurwitz, A. [1893] ``\"{U}ber die Bedingungen, unter welchen eine Gleichung nurWurzeln mit negativen reellen
Theilen besitzt,'' {\it Math. Ann.} 41, p.~403-442.

\bibitem[Klein \& Baier(1991)]{KleinBaier1991}
Klein, M. \&  Baier, G. [1991] {\it Hierarchies of dynamical systems}, In {\it A Chaotic Hierarchy}, edited by G. Baier and M. Klein. Singapore: World Scientific.

\bibitem[Kuznetsov(2004)]{Kuznetsov}
Kuznetsov, Yu. A. [2004] {\it Elements of Applied Bifurcation Theory}, Springer-Verlag, New York, third edition.

\bibitem[Lang \& Kobayashi(1980)]{Lang}
Lang, R. \& Kobayashi, K. [1980] ``External optical feedback effects on semiconductor injection laser properties,'' \emph{IEEE Journal of Quantum Electronics}, 16 (3) 347-355.


\bibitem[Lorenz(1963)]{lorenz1963}
Lorenz, E. N. [1963] ``Deterministic nonperiodic flow,'' \emph{J. Atm. Sci.} 20, 130-141.

\bibitem[Lotka(1910)]{lotka1910}
Lotka, A. J. [1910] ``Contribution to the theory of periodic reaction,'' \emph{J. Phys. Chem.} 14 (3) 271-274.

\bibitem[Lotka(1920)]{lotka1920}
Lotka, A. J. [1920] ``Analytical note on certain rhythmic relations in organic systems,'' \emph{Proc. Natl. Acad. Sci. U.S.A.}, 6(7) 410-415.

\bibitem[Marsden \& McCracken(1976)]{Marsden}
Marsden, J. \&  McCracken, M. [1976] {\it Hopf Bifurcation and its Applications}, Springer-Verlag, New York.

\bibitem[Matsumoto(1984)]{matsumoto1984}
Matsumoto, T. [1984] ``A chaotic attractor from Chua's circuit,'' \emph{IEEE Transactions on Circuits and Systems}, 31, 1055-1058.

\bibitem[Ohtsubo \& Davis(2005)]{Ohtsubo}
Ohtsubo, J. \& Davis, P. [2005] ``Chaotic optical communication,'' In: Kane D, Shore KA (eds) \emph{Unlocking dynamical diversity-optical feedback effects on semiconductor lasers}, Chap. 10. Wiley, Chichester.


\bibitem[Pecora \& Carroll(1990)]{pecoracarroll1990}
Pecora, L. M. \& Carroll, T. L. [1990] ``Synchronization in chaotic systems,'' \emph{Phys. Rev. Lett.} 64, 821-824.

\bibitem[Roessler(1976)]{roessler1976}
Roessler, O. E. [1976] ``An equation for chaos,'' \emph{Phys. Lett.}, 57A (5), 397-398.

\bibitem[Routh(1877)]{Routh1877}
Routh, E. J. [1877] \textit{A Treatise on the Stability of a Given State of Motion: Particularly Steady Motion}, Macmillan and co.

\bibitem[Sandri(1996)]{sandri1996}
Sandri, M. [1996] ``Numerical calculation of Lyapunov exponents,'' \emph{The Mathematica Journal}, \textbf{6}(3), 78-84.

\bibitem[Schwartz \& Smith(1983)]{SEIR}
Schwartz, I. B.  \& Smith, H. [1983] ``Infinite subharmonic bifurcation in an SEIR epidemic model,'' \emph{Journal of mathematical biology}, 18(3) 233-253.

\bibitem[Sprott(2003)]{sprott}
Sprott, J. C. [2003] \emph{Chaos and Time-Series Analysis}, Oxford University Press.

\bibitem[Van Wiggeren \& Roy(1998)]{roy}
Van Wiggeren, G. D. \&  Roy, R. [1998] ``Communication with chaotic lasers,'' \emph{Science}, 279, 1198-1200.

\bibitem[Varone {\it et al.}(1995)]{varone1995}
Varone, A., Politi, A. \& Ciofini, M. [1995]  ``$CO_2$ laser dynamics with feedback,'' \emph{Phys. Rev. A} 52, 3176-3182.

\bibitem[Volterra(1926)]{volterra1926}
Volterra, V. [1926] ``Variazioni e fluttuazioni del numero d'individui in specie animali conviventi,'' \emph{Mem. Acad. Lincei Roma}, 2 31-113.

\bibitem[Volterra(1931)]{volterra1931}
Volterra, V. [1931] ``Variations and fluctuations of the number of individuals in animal species living together,'' In: Chapman, R.N., Ed., \emph{Animal Ecology}, McGraw-Hill, New York, 409-448.

\bibitem[Weiss \& Brock(1986)]{PRL1986}
Weiss, C. O. \&  Brock, J. [1986] ``Evidence of Lorenz-type chaos  in a laser,'' \emph{Phys. Rev. Lett.} 57, 2804-2806.

\bibitem[Wolf {\it et al.}(1985)]{wolf1985}
A. Wolf, J.B. Swift, H.L. Swinney \& J.A. Vastano, Determining Lyapunov exponents from a time series, \emph{Physica D}, \textbf{16}, 285-317 (1985).


\end{thebibliography}
\end{document}